\pdfoutput=1
\documentclass[a4paper,10pt]{article}
\usepackage[english]{babel}
\usepackage{amsmath}
\usepackage{graphicx}
\usepackage{graphics}
\usepackage{multirow}
\usepackage[tiny]{titlesec}
\title{a first thermodynamic interpretation of the technology transfer activities}
\tiny{\author{s.ripandelli and u.lucia (Politecnico di Torino)}}
\begin{document}
\maketitle
\section*{Abstract} In the last years new interdisciplinary approaches to economics and social science have been developed. A Thermodynamic approach to socio-economics has brought to a new interdisciplinary scientific field called econophysics. Why thermodynamic? Thermodynamic is a statistical theory for large atomic system under constraints of energy[1] and the economy can be considered a large system governed by complex rules. The present job proposes a new application, starting from econophysic, passing throughout the thermodynamic laws to interpret and to described the Technology Transfer (TT) activities. Using the definition of economy (i.e. economy[dictionary def.] = the process or system by which goods and services are produced, sold, and bought in a country or region) the TT can be considered an important sub-domain of the economy and a transversal new area of the scientific research. The TT is the process of transferring knowledge, that uses the results from the research to produce innovation and to ensure that scientific and technological developments could become accessible to a wider range of users. Starting from important Universities (MIT, Stanford, Oxford, etc) nowadays the TT is assuming a central role. It is called the third mission of the research, together with education and research. The importance to provide new theories and tools to describe the TT activities and their behavior, has been retained fundamental to support the social rapid evolution that is involving the TT offices. The presented work uses the thermodynamic theories applying them to Technology Transfer and starting from the concept of entropy, exergy and anergy. The analysis output should become an help to make decision to improve the TT activities and a better resources employment.
\section*{General introduction}
Before to propose a thermodynamic model of the TT activities, it was necessary to identify in which environment the TT offices operates, schematizing the boundary conditions that, such for all thermodynamics complex systems, influence the modification of the system \cite{1,2}. In figure 1, a general representation has been provided in order to share with the reader a common starting point, necessary to build on the theory \cite{3}.
\begin{figure}[htb]
\centering
\includegraphics[scale=0.3]{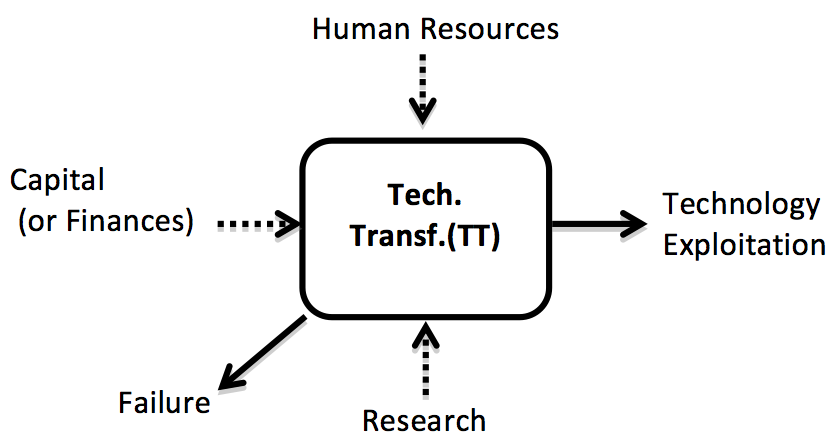}
\caption{thermodynamic interpretation of the activity of TT}
\end{figure}
From figure 1 a more general scheme, presented in figure 2, shows two important aspects of the theory. When a new technology is transferred, a general transformation occurs (i.e. changing on Innovation Level). 
\begin{figure}[htb]
\centering
\includegraphics[scale=0.3]{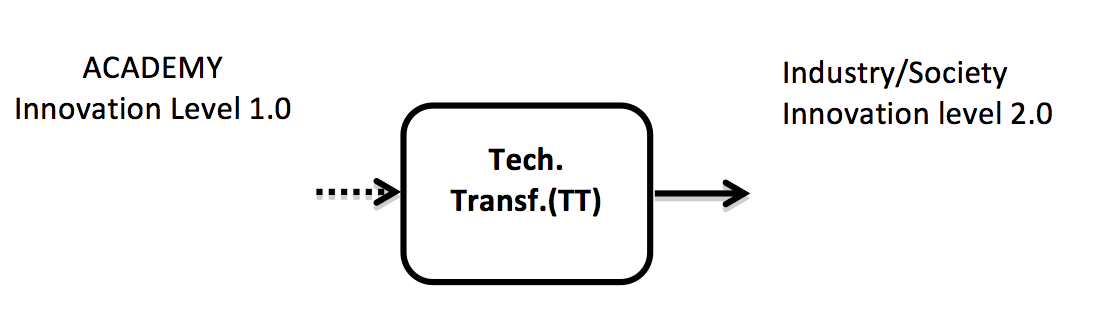}
\caption{a simplified scheme of the activity of TT}
\end{figure}
The TT supports this transformation, passing from an idea to “laboratory” to the “real world”. The transformation regards an implementation of the technology, and qunatitatively is estimated with the help of the Technological Readiness Level scale\cite{4,5,6,7}. From another point of view, using a chemical reaction analogy, the TT operates such as a reaction catalyst and promotes the transformation, trying to guide the technology implementation on the right direction.
Starting from this idea a generic three steps based transformation occurs periodically.
1. A first instability step is generated from a new discover or innovative research;
2. a development step: applied research, experiments and final prototypes;
3. the new technology reaches the market, influencing it: a new equilibrium will be reached.
When a system reachs an equilibrium, if this is not an absolute equilibrium, a new transformation can occur under specific new conditions. It is the case of the innovation. Once an innovation meets the real world it will modify it, and after a first period of instability a long period of stability occurs. Describing the system at time t$_0$ with a proper energy surface, when a modification occurs even the associated energy surface will be modified. The transformation, under certain conditions, shall modify the energy surface and the system shall reach a new local minimum. For each transformation a new analysis is needed, in order to analyze correctly its evolution over time.
\section*{The thermodynamic formulation}
Such as for the classical thermodynamics it was necessary to analyze the nature of the variables. They can have double nature. Intensive variable are independent from the amount of the species that are part of the system, while the extensive variables depend on the variation on the dimension of the system. A thermodynamic system is described by a mix of these variables.
In order to describe the TT activity has been decided to start from the thermodynamic equation, or the free Gibbs equivalent energy equation \cite{8,9,10,11}:
\begin{equation}
\small{
G=U+pV-TS=-U_p+cT+pV-TS 
}
\end{equation}
In thermodynamic this equation makes an analysis on the internal energy, the modification of the system due to external influence and involves even the concept of entropy. The entropy variation describes the modification of the system that occurs when each system is evolving.
The equation 1 has been considered a good and innovative starting point able to give to TT officers an instrument to evaluate rapidly how and if the system is working in a right direction. Obviously this analysis cannot be exhaustive, because the environment in which the TT operates is constantly in evolution. Anyway, I retain that this so complicated system can be simplified as a sum of discretizes sub-systems: at each step an equilibrium is reached, in accord with the periodic transformation described into the previous chapter.
In order to employ the equation 1, it was necessary to propose a variables transposition, even identifying which are the intensive and the extensive variables from the TT point of view. Here it is the table 1, reporting all the variables such as were thought. \\
\begin{table}[htbp]\centering 
\begin{tabular}{l c c  c}\hline\hline
\multicolumn{1}{c}{\textbf{Variable}} & \textbf{-}
 & \textbf{Thermodynamic} & \textbf{TT equivalent}\\ \hline
Temp.(T) & Intensive & vibration & economical activity\\
Const.(c) & Intensive & -  & Num. activities\\
Entropy(S) & Intensive & Syst. Evolution  & TT pro-activity\\
Volume(V) & Exstensive & -  & human resources in euro\\
Pressure(P) & Intensive & -  & external to internal influences\\
Int.En.(Up) & Exstentive & - & potential at starting point\\
Gibbs(G) & - & =U+pV-TS  & =-Up+cT+pV-TS\\
\hline\end{tabular}
\caption{the equivalence with the thermodynamics variables}
\end{table}
Once to have discuss about the nature of the variables, it was necessary to understand in which way was possible to use the raw data into the formulation proposed, and in which way the formulation can provide a useful result.
The approach employed is the sequent. A time interval has been chosen. The first assumption was that in that interval the external conditions stay unmodified. At time t0 the internal system (hereafter called TTs) can received external inputs. From time t0 to t1 the TTs operates and at t1 its activity has been analyzed to compute the goodness of its activities. At this step it is important to underline how not all the activities carried out by the TT have immediately an impact (positive or negative as well). For this reason it is important to operate in two directions:
\begin{itemize}
\item[-] to carry out an analysis in order to provide an quantitative analysis of the TTs, giving a feedback on the exploitation of the invested resources and suggesting in which sectors much more efforts are necessary;
\item[-] to collect periodically data to ensure a constant update on the status of the TTs helping its management.
\end{itemize}
The values that eq.1 assumes is able to give immediately a measure of the TTs quality and how it is working. Anyway it is important to underline how the approach proposed considers the TT office and its activity is part of a composed thermodynamic system where the Innovation is the environment where it operates. 
The scheme proposed in figure 2 it is fundamental to justify the thermodynamic approach. It was not possible to analyze the TT activity reducing it as a simple system. Its activity, indeed, is only a part of the Innovation ecosystem. Moreover TT offices work promoting open collaboration with other systems. This activity modifies periodically the boundary conditions and guide the E$_{TT}$ system through a new equilibrium state passing from non-equilibrium ones.
From a theoretical point of view every sub-system, part of a composed bigger system, can be describe with a succession of single quasi-static transformations. When a condition (or more than one) changes the system flows looking for a new equilibrium state and, after a period τ, that we call transition time, reaches a new equilibrium waiting for a new change.
The equation 1 can be assume three different value:
\begin{itemize}
\item[$<$]0;  for an evolving status throughout an higher efficiency. The boundary conditions guide the system in a spontaneous transformation. The entropy value is, by definition, increasing.
\item[=]0;  the system has a limit condition. No irreversible transformation occurs. This condition represents a boundary between a positive an negative exploitation of resources. 
\item[$>$]; the system transformation is not spontaneous. 
\end{itemize}
\section*{Example of calculus for the model proposed}
In order to show in which way the model works, an example was performed using the data from the TT Lab of Politecnico di Torino. A particular analysis has been conducted using data from patents and the activities carried out by internal Project Managers. From January 2015, the TT area of Politecnico di Torino has founded an interdisciplinary laboratory (LabTT) in order to create a team having transversal competences in different scientific field, trying to help researchers on the TT activities. Politecnico di Torino invested some money to promote this initiative. Starting from the investments done by Politecnico di Torino, has been computed the positive effects comparing the results to previous data when the LabTT was not constituted yet.
Analyzing the equation (1) each term has been described by the following equations.
Up is the potential energy of the TT. Up is the potential internal energy and represents the money invested into the technology transfer at the start of each period.
\begin{equation}
\small{
∆E_p=\int_{0}^{t}\frac{dE_p}{dt} dt= 〖-U〗_p^{ij}+\sum_j▒〖[(c_{Lj}-c_j ) T_0-c_j (T_{Rj}-Q_j^{(IN-OUT)} )]       (E_p  ≡ U_p^{ij}+c_j T)〗
}
\end{equation}
When anybody exploit every kind of resources and a transformation of them occur, the thermodynamic tell us that the entropy changes. A transformation generates a variation of entropy. If dS = 0 the transformation is reversible while, for all real transformation dS$>$0.
So it is easy to obtain the following equation.
\begin{equation}
\small{
∆S=\int_{0}^{t}\frac{dS}{dt}dt=\sum_m▒〖[(-S_{Gm}+S_{Lm} )+S_{gm}+F_m ]∙〗 T_m
}	
\end{equation}
The entropy variation in the equation 3 shows the presence of four components. S$_{Gm}$, S$_{Lm}$, S$_{gm}$ and F$_{m}$ represent the components of entropy that have generated positive results, for which one no results have been obtained, the entropy from energy lost and that one necessary for the LabTT to carry out its activities respectively. These two factor Sgm and Fm correspond to values that described the irreversible nature of the TT activity. Without these two factors, each transformation that occurs while a technology is “incubated” to make innovation, wouldn’t constitute a challenge.
Posing the value of Sgm and Fm equal to 0 the entropy variation for reversible transformation is obtained. The ratio between the value obtained from equation (1), or associated to irreversible and real events, and the value of Gibbs equivalent posing S$_{gm}$=F$_{m}$=0 is that one employed in this study to do an evaluation about the activity of TT Lab of Politecnico di Torino.
\begin{table}
\begin{tabular}{|c|c|c|c|c|c|c|}
\hline
\multicolumn{7}{|c|}{\textbf{2015-2016}}\\
\hline
\multicolumn{1}{|c|}{\textbf{Variables}} & \textbf{Description} & \textbf{[-]} & \multicolumn{4}{c}{\textbf{Time}}\\
\hline
& & &2015I&2015II&2016I&2016II\\
\hline
\multirow{2}{*}{c$_d$}  & \multirow{2}{*}{Num. of Disclosures} & - & 40 & 30 & 30 & NA \\ 
 & &keuro & 0 & 0 & 0 & NA\\
\hline
\multirow{2}{*}{c$_f$}  & \multirow{2}{*}{Num. of Filed Patents} & - & 15 & 6 & 10 & NA \\ 
 & &keuro per patent & 2.0 & 2.0 & 2.0 & 2.0\\
\hline
\multirow{2}{*}{T$_d$}  & \multirow{2}{*}{LabTT Officer} & person & 5 & 4 & 3 & NA \\ 
 & &keuro/person & 20 & 20 & 20 & 20\\
\hline
\multirow{2}{*}{T$_b$}  & \multirow{2}{*}{Others} & person & 3 & 5 & 8 & NA \\ 
 & &keuro/patent & 22 & 22 & 22 & 22\\ 
\hline
{S$'_c$}  & Rev.to TT activities & keuro & 30 & 10 & NO & NA \\ 
\hline
{Q$^{In-Out}$}  & Profit & keuro & 25 & 0 & 0 & NA \\
\hline
{S$'_i$}  & Other & keuro & 0 & 0 & 0 & NA \\
\hline
{S$_{TRL}$}  & Work on TRL & keuro & NA & NA & NA & NA \\
\hline
\end{tabular}
\caption{the equivalence with the thermodynamics variables}
\end{table}
For each m project it is possible to identify these four parameters, and for every project is known previously the amount of money invested (T$_{m}$).In the following table the variables are 
Using data from table 1 the equation 1 can be written as follow:
Numerical Example, related to the patent activities of Politecnico di Torino (all data employed are fictitious only to show how the theory works) has been proposed.
\begin{equation}
\small{
\begin{split}
∆G=-U_p^{ij}-\sum_{j}[(c_dj-c_fj ) T_p-c_{fj} (T_p-Q_j^{IN-OUT} )]+\\-\sum_{k}▒〖[(-〖S_{Gk}〗^{'}-〖S_{ik}〗^{'} )]-〗 \sum_{m}▒S_{Lm}  T_cf-\sum_{n}▒[(S_{gn}+F_n )∙T_d ]  
\end{split} 
}
\end{equation}
The members assume the values reported in table 2. 
For the first period analyzed, 2015I, the values assumed the following values. Up = 70 [keuro]. T$_0$ represents the amount of resources, expressed in euro, required to carry out the activities related to the patents. For the specific example reported are:
\small{
\begin{gather*}
T_0= T_p=\frac{(2.0*20*5)}{40}=5.0 \quad [keuro];\\
Q_j^{(IN-OUT)}= 25 \quad [keuro];\\
T_{cf}=2.2 \quad [keuro]; \\ \\
T_Td=20 \quad [keuro];  \\
F_n = T_d=5 \quad [people \quad in \quad LabTT];\\
S_{gn}=  \frac{(25 *10 )}{182.5}= 1.37 \quad [keuro];\\
〖S'〗_{Gk}=S_{Gk}∙T_cf= 30 \quad [keuro \quad equivalent \quad gained]; \\
S_{Lm}=\frac{(15*10)}{182.5}=0.83 \quad [keuro \quad equivalent \quad lost];  
\end{gather*}}
\normalsize{
Replacing these values into the equation (4) is possible to obtain $\Delta$ G$^I$= - 344.226 [keuro]. The equation (4) is also computed setting the parameter  S$_{gn}$ = S$_{Gm}$ = F$_{n}$=0 , obtaining the value of  $\Delta$G$_g^I$= -246.826 [keuro]. This last value represents the  Gibbs energy equivalent for a system able to exploit all the internal energy without losses. In a thermodynamic sense the maximum of energy is equal to the all internal energy, represents by all resources employed, transformed in positive results.  In other words the total anergy, that for all real transformation never assumes null values, is equal to 0.
Computing the ratio between the two values obtained, $\Delta$G$_{g}$/$\Delta$G, can assumes three values:
\begin{itemize}
\item[-]	 $\Delta$G$_{g}$/$\Delta$G$<$1 ; The system does not exploit well the resources 
\item[-]	$ \Delta$G$_{g}$/$\Delta$G$=$1 ; the TT activitiy appears as the counterpart of a complete reversible phenomena in thermodynamics 
\item[-]	 $\Delta$G$_{g}$/$\Delta$G$>$1 ; the results assumes a “super-reversible” behavior. The resources are exploit well.
\end{itemize}
For this specific case  the ratio results have been reported in table 3 (initial internal energy was considered always equal to 70keuro). Talking about the thermodynamics transformations, when $\Delta$G$_{g}$ is $<$0 a reaction is spontaneous. A comparison between $\Delta$G$_{g}$ and $\Delta$G put in evidence when the ratio is higher than one and when it has a positive value. The activities carried out exploit better the resources reaching a value $0<$$\Delta$G$_{g}$/$\Delta$G$<$1. In particular higher is delta from 1, higher is the distance from optimum.
\begin{table}[htbp]\centering 
\begin{tabular}{|l| c| c| c|}\hline
\multicolumn{1}{|c|}{\textbf{Time}} & \multicolumn{3}{|c|}{\textbf{Gibbs Energy}}\\ \hline
 & $\Delta$G [keuro] & $\Delta$G$_g$[keuro]  & $\Delta$G$_{g}$/$\Delta$G[keuro] \\ \hline
2015I & -594.2	&-262.96	& 0.83\\
2015II & -262.96	& -166.66 &	0.634\\
2016I & -193.123 &	-111.2	 & 0.576\\
2016II & NA & NA & NA\\
\hline\end{tabular}
\caption{the method applied on the three periods identified}
\end{table}
\section*{Conclusion}
The study has been conducted in order to understand, after a detailed analysis of all activities carry out by the TT offices, in which direction the Technology Transfer is going on. In the next future a quantitative and complementary analysis, following this study, would offer an evaluation tool to support decisional aspects. 
It is important to underline how the approach here proposed does not have the aim to provide a forecast or an exhaustive response about the quality of the work carried out by TT offices. This is impossible, indeed, for all dynamic non-linear systems. At this stage the method tries to take a picture of a dynamic system, in a range of time previously defined. 
In the next future more detailed analysis will be investigated, expanding the variables analyzed and applying the method to the different activities that involved a TT office in everyday life.

\end{document}